\documentclass[a4paper]{jpconf}
\usepackage{graphicx}
\usepackage[numbers,square,sort&compress]{natbib}
\usepackage{amsmath}
\usepackage{amssymb}
\usepackage{aas_macros}

\newcommand{\rp}{r_{\rm p}}
\newcommand{\wc}{w_{\rm c}}
\newcommand{\au}{{\, \rm au}}

\begin{document}


\title{Low mass planet migration in Hall-affected disks}

\author{Colin P McNally,$^{1}$
Richard P Nelson,$^{1}$
Sijme-Jan Paardekooper,$^{1,3}$
Oliver Gressel $^{4}$
and Wladimir Lyra $^{5,6}$}

\address{
$^{1}$ Astronomy Unit, School of Physics and Astronomy, Queen Mary University of London, London E1 4NS, UK}
\address{
$^{3}$ DAMTP, University of Cambridge, Wilberforce Road, Cambridge CB3 0WA, UK}
\address{
$^{4}$ Niels Bohr International Academy, The Niels Bohr Institute, Blegdamsvej 17, DK-2100, Copenhagen \O, Denmark}
\address{
$^{5}$ Department of Physics and Astronomy, California State University Northridge, 18111 Nordhoff St, Northridge, CA 91330, USA}
\address{
$^{6}$ Jet Propulsion Laboratory, California Institute of Technology, 4800 Oak Grove Drive, Pasadena, CA 91109, USA
}

\ead{c.mcnally@qmul.ac.uk}

\begin{abstract}
Recent developments in non-ideal magnetohydrodynamic simulations of protoplanetary disks
suggest that instead of being traditional turbulent (viscous) accretion disks,
they have a largely laminar flow with accretion driven by large-scale wind torques.
These disks are possibly threaded by Hall-effect generated large-scale horizontal magnetic fields.
We have examined 
the dynamics of the corotation region of a low mass planet embedded in such a disk and the evolution of the associated migration torque.
These disks lack strong turbulence and associated turbulent diffusion, and the presence of a magnetic field and radial gas flow 
presents a situation outside the applicability of previous corotation torque theory.
We summarize the analytical analysis of the corotation torque, 
 give details on the numerical methods used, and in particular the relative merits of different numerical schemes for the inviscid problem.
\end{abstract}

\section{Introduction}
\begin{figure}[h]
\begin{center}
\includegraphics[width=0.5\textwidth]{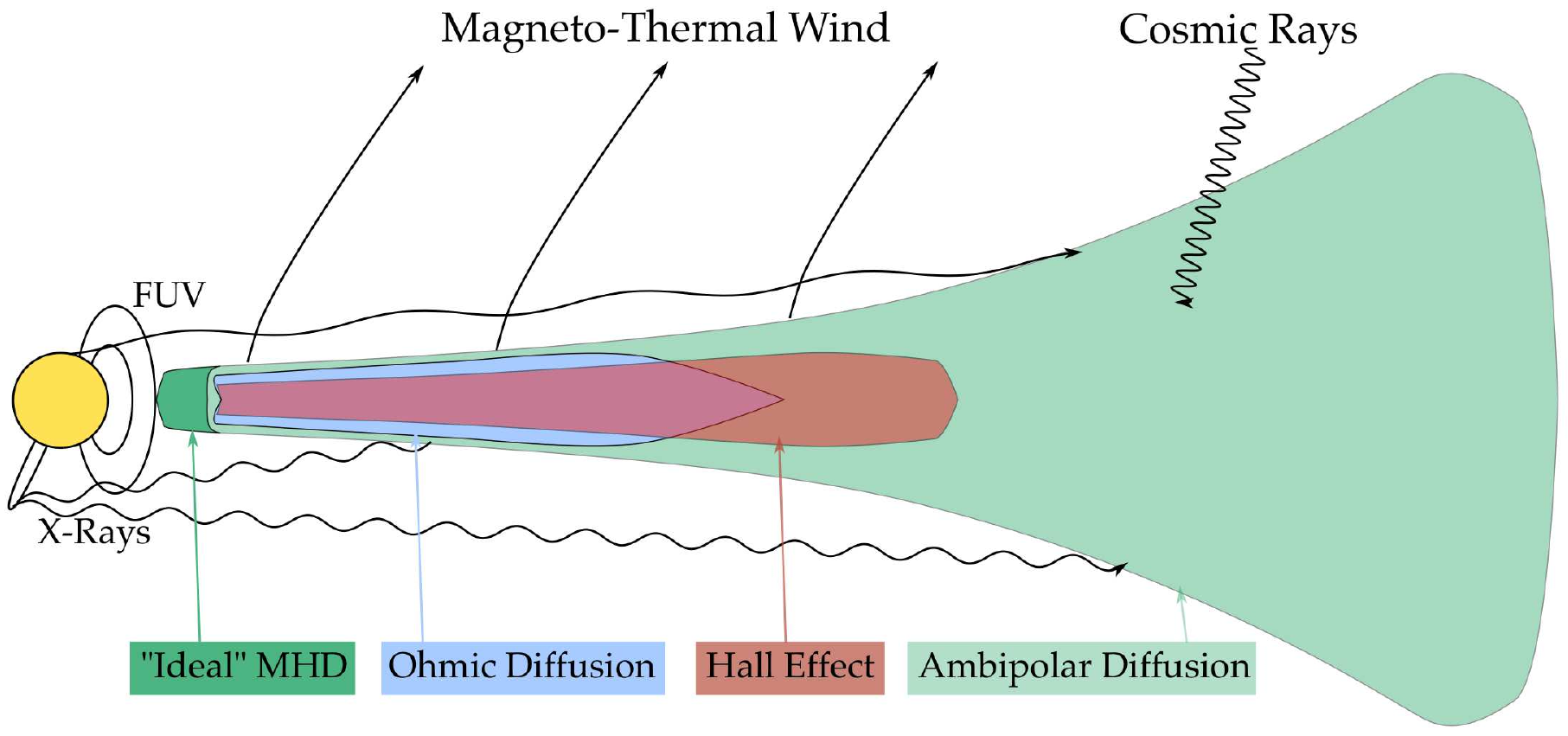}
\quad
 \begin{minipage}[b]{2.5in}
\caption{Schematic of the structure of a wind-driven protoplanetary disk in terms of ionization sources and non-ideal MHD terms.
}
\end{minipage}
\end{center}
\label{fig:schematic}
\end{figure}

Protoplanetary disks, the formation sites of solar systems, are comprised of a very low ionization gas and a mixture of dust, boulders, and protoplanets.
Simulations have shown that if the disk is threaded by a weak net vertical field, 
the disk develops an essentially laminar flow, as the magnetorotational instability is quenched by large Ohmic and Ambipolar diffusion.
The spatial arrangement of these regimes is  sketched in  Figure~\ref{fig:schematic}.
The accretion is  driven by a wind from the most ionized region near the disk surfaces \citep{2013ApJ...769...76B,2015ApJ...801...84G}.
Near the disk midplane, where Ohmic diffusion dominates, a `dead zone' forms ($0.5 \lesssim R \lesssim 10 \au$ from the star),
but at the same time the large Hall effect can give rise to strong fields when 
the background field is aligned with the disk rotation \citep{1999MNRAS.303..239W,2008MNRAS.385.2269P}.
These laminar magnetic fields can give rise to significant accretion at the midplane \citep{2013ApJ...769...76B,2014A&A...566A..56L,2017A&A...600A..75B}. 
Recently in \citep{2017MNRAS.472.1565M} we considered the problem of low mass planet migration in such a wind-driven protoplanetary disk 
that support accretion at the midplane
\citep{
2013ApJ...769...76B,
2013ApJ...772...96B,
2014ApJ...791...73B,
2014ApJ...791..137B,
2014A&A...566A..56L,
2015ApJ...798...84B,
2015MNRAS.454.1117S,
2016ApJ...818..152B,
2016ApJ...821...80B,
2017A&A...600A..75B}. 
In this proceeding we summarize the problem and discuss the numerical issues in greater detail.

To study the flow near a low mass planet, we adopt a common two-dimensional formulation of the disk-planet interaction, and reduce the Hall-generated field to this dimensionality. In  \citep{2017MNRAS.472.1565M}  we find a equilibrium field for the unperturbed disk:
\begin{align}
B_r = B_0 \left(\frac{r}{r_0}\right)^{-1} \quad B_\phi = -2 B_0 \Omega_0 r_0^2 \frac{\mu_0}{\eta} \left(\frac{r}{r_0}\right)^{-1/2} \label{eq:bphiic}
\end{align}
where $B_r$ and $B_\phi$ are the radial and azimuthal magnetic fields, $r$ is the cylindrical radius, $r_0$ is a fiducial radius, 
$\Omega_0$ is the Keplerian orbital frequency at $r_0$, $\eta$ is the Ohmic resistivity, and $\mu_0$ the vacuum permeability.
This magnetic field drives radial motion of the disk gas at a velocity denoted $v_r$.
Importantly, the resistivity of the dead zone region we study is very high, such that the timescale for magnetic field perturbations to diffuse across the length scale of interest (the planet's corotation region) is very small compared to the timescale for fluid elements to make the horseshoe turn 
at the end of a librating orbit in that corotation region. So, although the disk is well enough ionized so that the magnetic field drives the radial motion of the disk gas,
the magnetic field response to the presence of the planet may be very small. 
Indeed, upon investigation we found it to to be negligible.

\section{Corotation Torques}

\begin{figure}
\begin{center}
\includegraphics[width=0.3\textwidth]{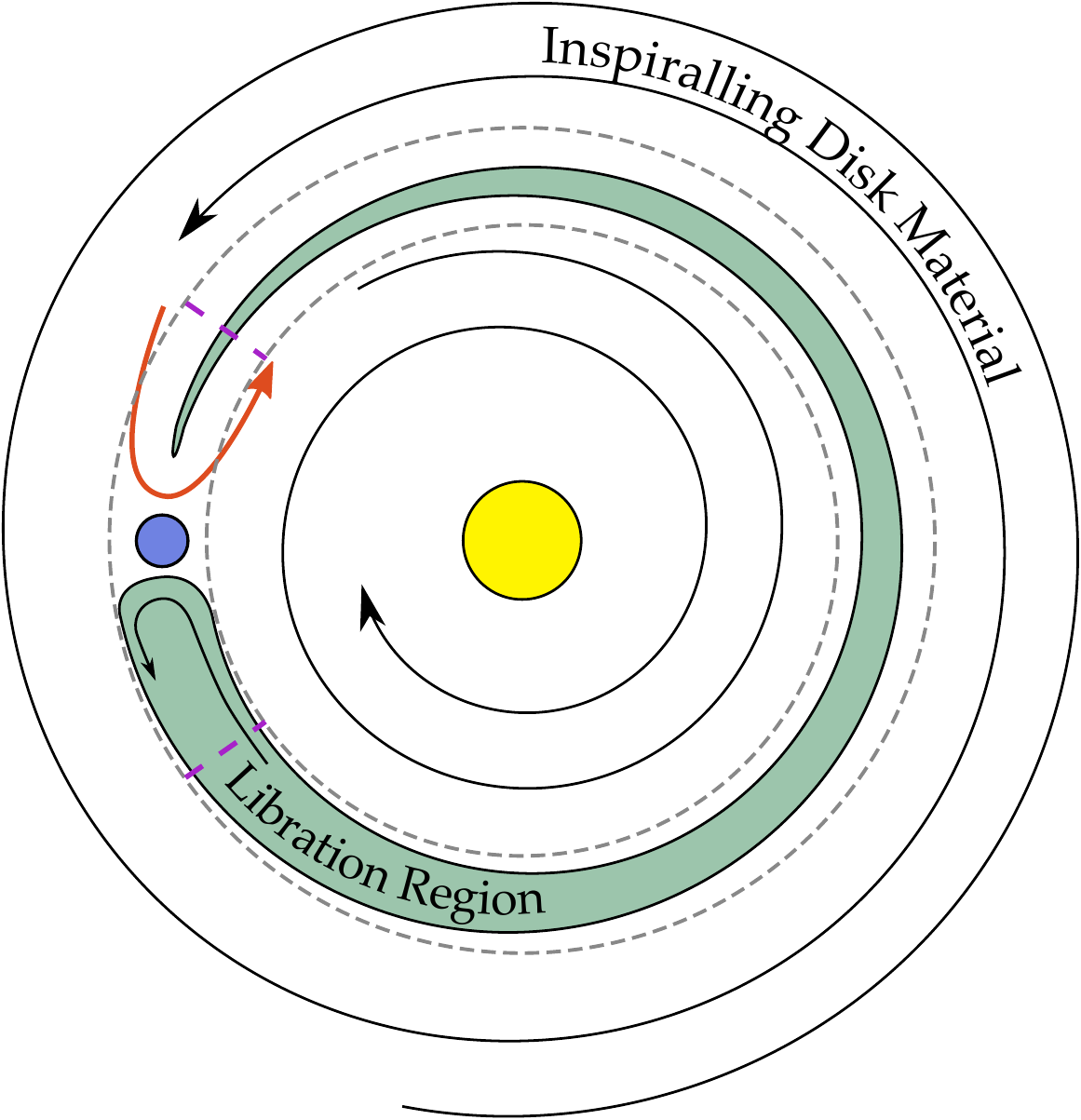}
\quad
 \begin{minipage}[b]{3.5in}
\caption{Cartoon of the flow in and around the corotation region of a low mass planet in a magnetically torqued dead zone.
The magnetic torque drives disk gas inwards, and distorts the island of librating material (green) into an asymmetrical tadpole-like shape.
This asymmetry of the librating material, and the introduction of the flow-through stream (red) lead to the corotation torque.}
\end{minipage}
\end{center}
\label{fig:flowcartoon}
\end{figure}

When a low mass planet is placed in a disk, the response of the disk to this perturbation creates under- and over-densities. 
These density perturbations gravitationally tug on the planet in the form of a torque which modifies its orbit.
Here, we focus exclusively on the corotation torque, arising from perturbations in the coorbital region, 
and for brevity neglect discussion of the spiral wave and associated Lindblad torque as they are not modified by the magnetic field in this case.
In this section we summarize the theory for the corotation torque arising in this inviscid flow by considering  the magnetic field as a fixed body force acting on the gas.
The presentation is a brief summary of the analytical results of \citep{2017MNRAS.472.1565M}.

In our disk with a laminar magnetic field driving accretion, the laminar flow past the planet is illustrated in Figure~\ref{fig:flowcartoon}.
Disk gas spiral inwards as it loses angular momentum to the magnetic field. When it encounters the planet from the outside 
it executes a turn, giving angular momentum to the planet and making a quick jump inwards toward the star. 
The width of the turn is $2 x_s$, where $x_s$ is the half-width of the corotation region.
Within the corotation region, there is gas that is trapped on librating orbits.
These orbits execute two horseshoe turns, in front and behind of the planet.
Due to the action of the magnetic field these orbits are also asymmetrical, as the gas drifts radially inwards between turns.
Thus, on the turn in front of the planet, both gas flowing through the coorbital region and gas on librating orbits makes the turn. 
In the turn behind the planet, only gas on librating orbits makes the turn.
The difference between the rate of angular momentum exchange with the planet by the gas on these two turns gives rise to the corotation torque.

To isolate the corotation torque due to the radial gas flow, and remove thermal effects (the entropy-related torque), we employ
a globally isothermal disk with  power law  surface density  $\Sigma = \Sigma_0 \left({r}/{r_0}\right)^{-\alpha}$.
In \citep{2017MNRAS.472.1565M} we show that the corotation torque can be calculated in terms of the inverse vortensity 
$w = \Sigma/\nabla \times \mathbf{v}$ of the material undergoing the two horseshoe turns.
Due to the steady-state magnetic accretion stress in the disk, as gas spirals inwards, the value of the vortensity at any given radial location
in the flow is constant, even though the vortensity of a given fluid element increases at it moves inwards.
At the same time, the gas in the libration region has constantly increasing vortensity because the same Lorentz force acts on it 
but it is not free to fall inwards towards the star.
The analysis in \citep{2017MNRAS.472.1565M} yields the expression for the torque associated with these two horseshoe turns
\begin{equation}
\Gamma_{\rm hs} = 2\pi \left( 1-\frac{\wc (t)}{w(\rp)}\right) \Sigma_{\rm p} \rp ^2 x_s \Omega_{\rm p} (-v_r) \label{eq:gammahsvr}\, ,
\end{equation}
where $w(\rp)$ is the inverse vortensity of the unperturbed disk at the planet location, $\wc(t)$ is 
the time-evolving characteristic value of the inverse vortensity in the libration region, $\Omega$ is the orbital frequency of the disk, $x_s$ is the corotation region half width, 
and quantities at the planet  position are denoted with a subscript $p$.
The corotation region half width for a low mass planet is approximately $x_s=1.2 \rp \sqrt(q/h)$
where $q$ is the planet-star mass ratio, and $h=H/r$ is the non-dimensional pressure scale height.
For the purposes of this discussion, we give the evolution of $w_c$ in a simple form as
\begin{align}
\wc(t) &= w(\rp) \left[ 1 + t/\tau_w \right]^{-1} \, ,   \label{eq:wcsol}\quad 
\tau_w  = \left[ \left(\frac{3}{2}-\alpha\right)\frac{( -v_r)}{ \rp} \left(\frac{\rp}{r_0}\right)^{\alpha-\frac{5}{2}} \right]^{-1}\, .
\end{align}
The full development of this formula, analysis of relevant timescales and its applicability are discussed in \citep{2017MNRAS.472.1565M}.

From this discussion of the analytical approach to the problem, we can see that to arrive at good numerically approximated solutions to this inviscid
 problem we require that the vortensity of the material trapped in the libration region should not artificially mix
with the flow-through material, so that the evolution of $w_c(t)$ is correctly obtained.
This becomes the main criteria for successfully running a fluid simulation.

\section{Methods for full magnetohydrodynamics}

The first code the problem was attacked with was {\sc NIRVANA3.5}
\citep{2004JCoPh.196..393Z,2017JPhCS.837a2008G}.
As employed here, it is a second order accurate Godunov method, with an HLLD Riemann solver and a 
CTU scheme for the magnetic field evolution and second-order Runge-Kutta time integration.
Because of the very large Ohmic diffusion, the parabolic term is  treated with a
Runge-Kutta-Legendre polynomial based second-order accurate
super-time-stepping  scheme (RKL2) \citep{2012MNRAS.422.2102M}, as previously employed in \citep{2015ApJ...801...84G}.
In addition, we have newly implemented a conservative
orbital advection scheme following \citep{2012A&A...545A.152M}, employing a FFT-based spectral interpolation for the azimuthal advection.

Typically, we find that the super-time-stepping allows a time step $\sim 2000$ times greater than the stability limit for a basic 
explicit finite difference integration of the diffusion operator, by employing a super time stepping solver with $\sim 65$ stages on 
each stage of the two-stage Runge-Kutta hydrodynamics time step with a wall clock speed-up of $\sim 17$ times over a simple explicit integration.
The RKL2 super time stepping scheme displayed no instability despite these extreme time step ratios.
Moreover, through these experiments 
we find that the problem can be described well by fixing the magnetic field and neglecting the induction equation.
This allowed the problem to be described as one in pure isothermal hydrodynamics, with a simple body force added to 
describe the Lorentz force from the magnetic field \citep{2017MNRAS.472.1565M}. 
Thus we were able to progress with a simplified problem, and search for even more efficient codes for this problem.

\section{The fixed magnetic field problem and vortensity evolution}
\begin{figure}
\begin{center}
\includegraphics[width=0.49\textwidth]{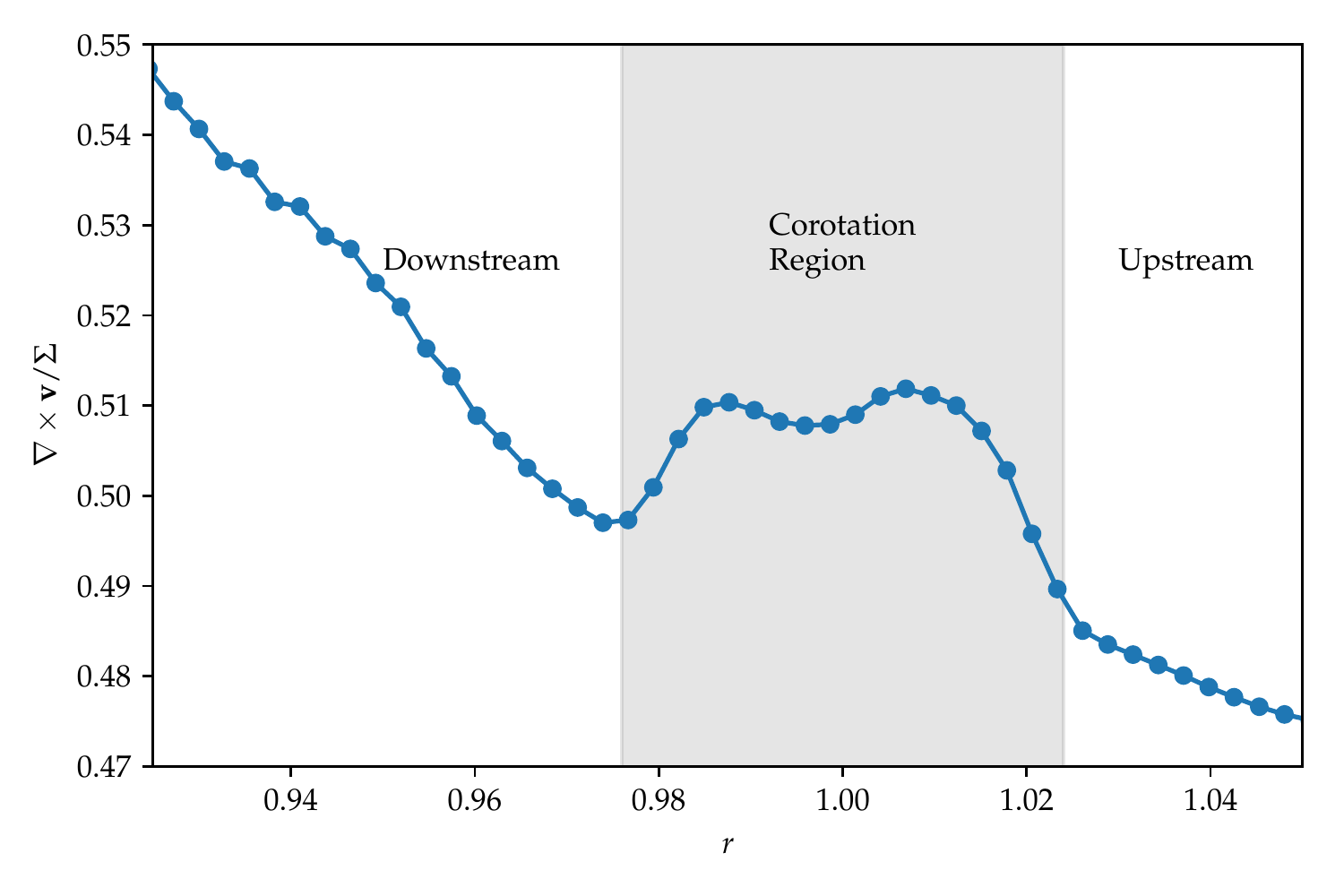}
\includegraphics[width=0.49\textwidth]{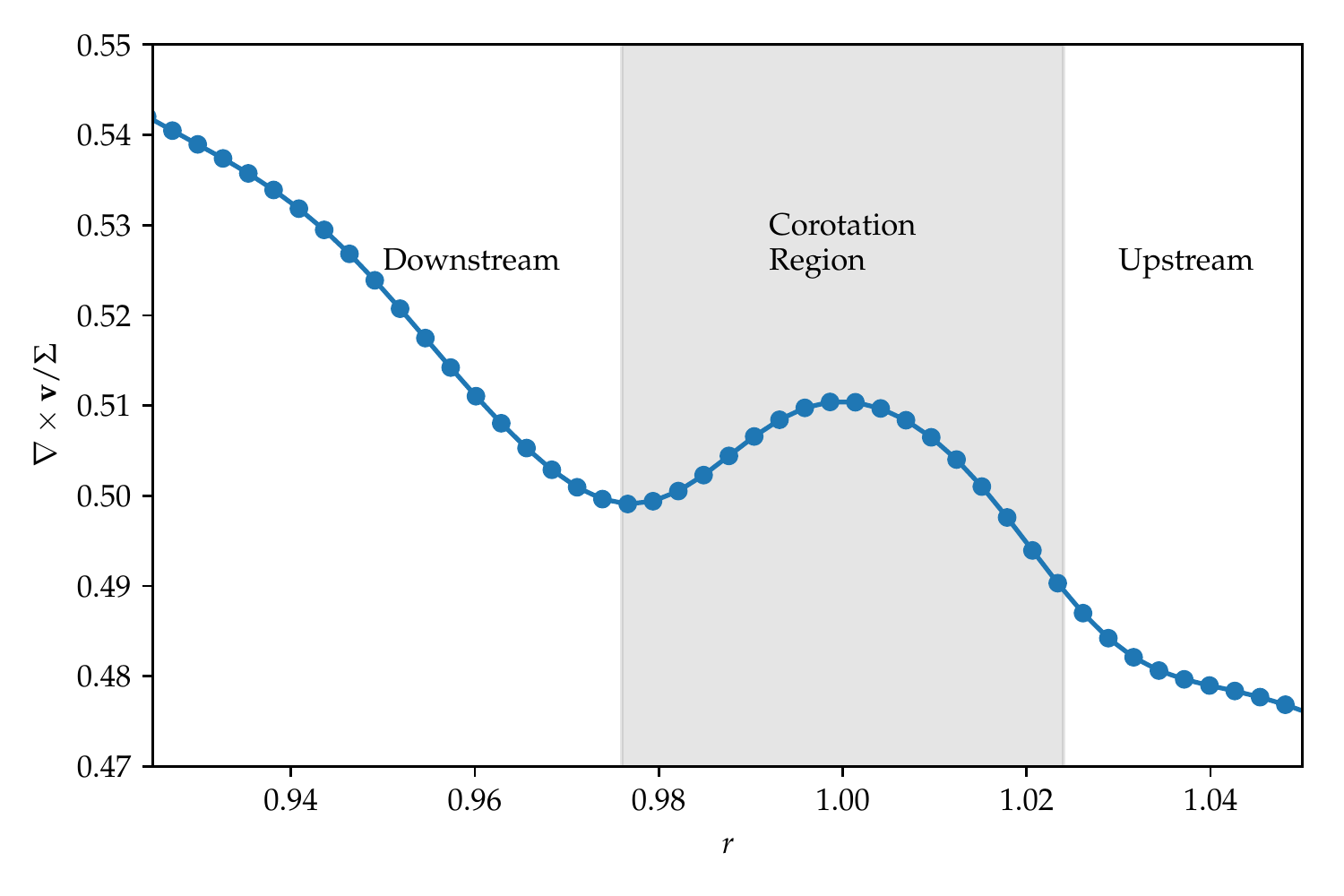}\\
\includegraphics[width=0.49\textwidth]{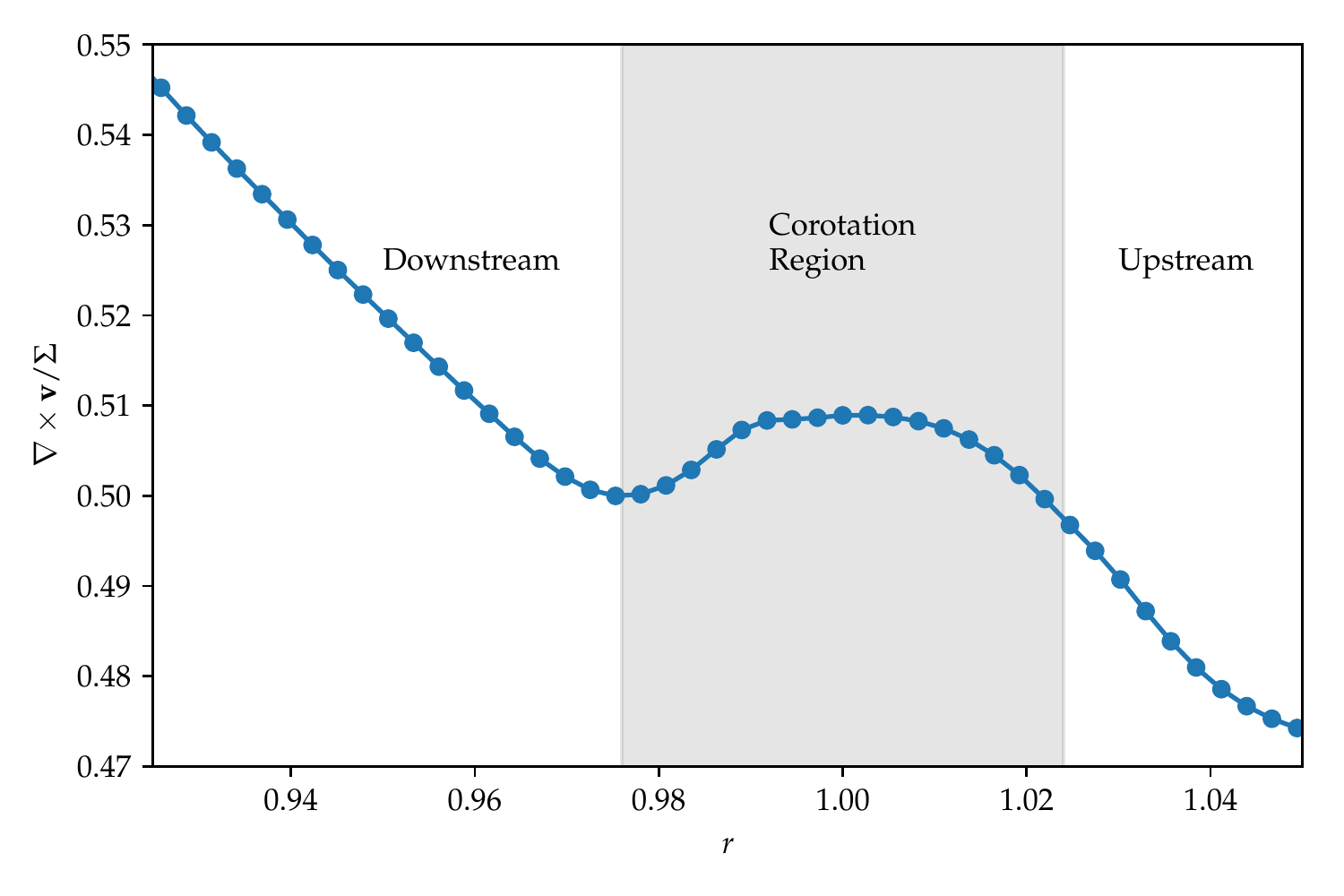}
\caption{Illustration of the difference between simulations in three codes. 
Examples of the vortensity perturbation in the corotation region, in a radial cut opposite to the planet after 240 orbits at a low resolution as specified in the text.
Upper Left: {\sc FARGO3D}, Upper Right: The {\sc Pencil Code}, Lower Center: {\sc NIRVANA3.5}}
\label{fig:vortcut}
\end{center}
\end{figure}

\begin{figure}
\begin{center}
\includegraphics[width=0.5\textwidth]{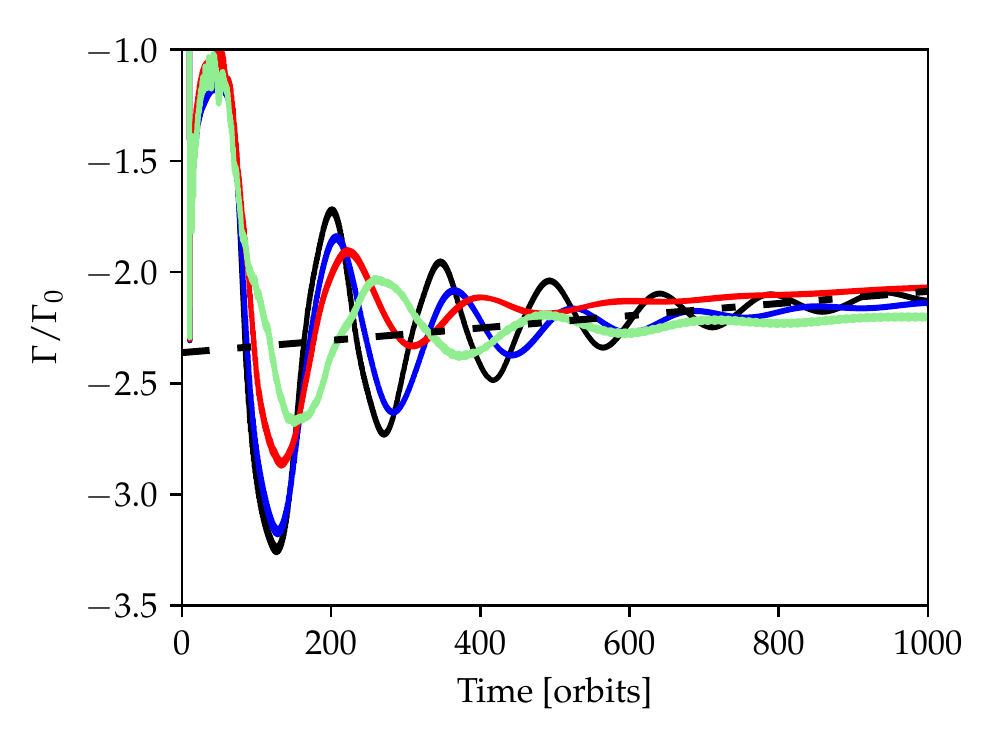}
\quad
 \begin{minipage}[b]{2.5in}
\caption{
Convergence of torque histories from {\sc FARGO3D} simulations and the analytical 
prediction.  Grid resolutions are doubled between each run, with the sequence light green, red, blue, black, and the analytical formula equation~(\ref{eq:gammahsvr}) is plotted with a dashed line.
 The total torque $\Gamma$ is scaled by  $\Gamma_0 = (q/h)^2 \Sigma_{\rm p} r_{\rm p}^4 \Omega_{\rm p}^2$.
}
\end{minipage}
\label{fig:torqueconv}
\end{center}
\end{figure}

The {\sc Pencil Code} \citep{2002CoPhC.147..471B,2010ascl.soft10060B} is a high-order finite difference method 
based on point collocation values and sixth-order centered differences in 
space to construct a method-of-lines approach to the MHD equations and applying a third-order time integration scheme.
In addition, for disk problems like the one here, we have implemented an orbital advection technique by
 splitting of the time integration  \cite{2017AJ....154..146L}, without decomposing the velocity field as in {\sc NIRVANA}. 
With this technique, the bulk azimuthal advection due to Keplerian motion is achieved with a FFT-based spectral interpolation, 
yielding a significant speedup and lower numerical diffusion while preserving the third order accuracy in time on the integration scheme.
Stability for the integration scheme is achieved by modifying the governing equations with the addition of diffusion operators.
The usual minimal choices in the {\sc Pencil Code} for a problem like the one at hand are a hyperdiffusion of the form $\nabla^6 = \nabla^2(\nabla^2(\nabla^2))$ 
applied to all fields to stabilize the grid scale, and a shock-viscosity. The
full form of the hyperdiffusion operator and the scaling of the hyperdiffusion coefficient 
as applied in cylindrical coordinates is given in the appendix of \cite{2017AJ....154..146L}.
Importantly, this hyperdiffusion has shear components which can mix vortensity into the libration region at the grid scale.

{\sc FARGO3D} \citep{2016ApJS..223...11B,2015ASPC..498..216M,2000A&AS..141..165M} is based on an operator splitting technique 
with a second-order conservative finite differences in space.
However, to gain stability the scheme employs only a bulk shock-viscosity with no shear components.
Additionally, the method is known to have good properties for the conservation and advection of vortensity ($\nabla \times \mathbf{v} /\Sigma$)
 \citep{2015ASPC..498..216M}.
Although the {\sc Pencil Code} has a high formal order of accuracy and has excellent fidelity for quickly evolving shear flow problems \citep{2012ApJS..201...18M}, 
for this problem the slow formation (hundreds of orbits) of the vortensity structure 
in the corotation region is washed out by the shear component of the stabilizing hyperdiffusion. 
Thus, although it has a lower formal order of accuracy, we found {\sc FARGO3D} to be a efficient in practice for achieving 
solutions with approximately the correct physical properties, particularly the sharp, relatively undiffused vortensity profile shown in Figure~\ref{fig:vortcut}.
This comparison uses a cylindrical grid of 512 zone in radius and 1536 in azimuth, with a radial range $[0.3,1.7]$ and boundary conditions
 and planet potential as employed in \cite{2017MNRAS.472.1565M}.
The planet/star mass ratio was $2\times10^{-5}$.
In the {\sc Pencil Code} the {\tt hyper3-cyl} \cite{2017AJ....154..146L} diffusions of velocity and density were employed with the  coefficient set to $10^{-4}$, 
the smallest value found to yield stability. In addition, the shock-capturing artificial viscosity {\tt nu-shock} was employed on the velocity field with the coefficient set to $1$. 
In this comparison, {\sc NIRVANA3.5} solved only hydrodynamics with the magnetic force replaced by a body force as in the other codes, 
and the monotonized central slope limiter was selected.
In agreement with previous commentary on Godunov methods  \citep{2015ASPC..498..216M} for similar problems, we found that 
the algorithm we applied in {\sc NIRVANA3.5} diffused the sharp vortensity profile in the corotation region more than {\sc FARGO3D}.
We note that the choice of limiter does have a significant effect on the  numerical diffusion of vortensity in that method.
With higher resolution versions of these {\sc FARGO3D} simulations, we were able to verify our analytical prediction for the corotation torque, 
as shown by the example in Figure~\ref{fig:torqueconv}.

\section{Conclusions}
In summary, we have:
\begin{itemize}
\item Implemented orbital advection schemes appropriate for disk problems in {\sc NIRVANA3.5} and the {\sc Pencil Code}
\item Demonstrated the practical application of RKL2 super time stepping  for Ohmic resistivity to very diffusive setups with no numerical stability issues arising
\item Described the relative strengths of the main methods used in {\sc NIRVANA3.5}, the {\sc Pencil Code}, and {\sc FARGO3D}, all well-proven and high-quality codes in their own rights, for an inviscid  disk problem where evolution 
of vortensity is paramount, finding the best results with {\sc FARGO3D}
\item Developed a torque formula for the corotation torque on a low mass planet embedded in a magnetically torqued dead zone
\end{itemize}

\ack
This research was supported by STFC Consolidated grants awarded to the QMUL Astronomy Unit 2015-2018  ST/M001202/1 and 2017-2020 ST/P000592/1. 
The simulations presented in this paper utilized Queen Mary's MidPlus computational facilities, supported by QMUL Research-IT and funded by EPSRC grant EP/K000128/1; the DiRAC Complexity system, operated by the University of Leicester IT Services, 
and the DiRAC Data Centric system at Durham University, 
operated by the Institute for Computational Cosmology
which form parts of the STFC DiRAC HPC Facility (www.dirac.ac.uk). 
This equipment is funded by BIS National E-Infrastructure capital grants ST/K000373/1, ST/K00042X/1; 
STFC capital  grant ST/K00087X/1, and STFC DiRAC Operations grants ST/K0003259/1 and ST/K003267/1. 
DiRAC is part of the National E-Infrastructure. 
This work used the {\sc NIRVANA3.5} code developed by Udo Ziegler at the Leibniz Institute for Astrophysics (AIP). This research was supported in part by the National Science Foundation under Grant No. NSF PHY11- 25915. 
The research leading to these results has received funding from the European Research Council (ERC) under the European UnionÕs Horizon 2020 research and innovation
programme (grant agreement No 638596). 
SJP is supported by a Royal Society University Research Fellowship.

\bibliography{astronum_mcnally_17.bib}

\end{document}